# A Hybrid Modified Semantic Matching Alogrithm Based on Instances Detection With Case Study on Renewable Energy

Ahmad Khader Haboush

Department of Computer Science, Jerash University

Jerash, Jordan

*Abstract*— This Matching input keywords with historical or information domain is an important point in modern computations in order to find the best match information domain for specific input queries. Matching algorithms represents hot area of researches in computer science and artificial intelligence. In the area of text matching, it is more reliable to study semantics of the pattern and query in terms of semantic matching. This paper improves the semantic matching results between input queries and information ontology domain. The contributed algorithm is a hybrid technique that is based on matching extracted instances from booth, the queries and in information domain. The instances extraction algorithm that is presented in this paper are contributed which is based on mathematical and statistical analysis of objects with respect to each other and also with respect to marked objects. The instances that are instances from the queries and information domain are subjected to semantic matching to find the best match, match percentage, and to improve the decision making process. An application case was studied in this paper which is related to renewable energy, where the input queries represents the customer requirements input and the knowledge domain is renewable energy vendors profiles. The comparison was made with most known recent matching researches.

*Keywords*— Semantic matching; Instances extraction; Decision making; Ontology; Renewable energy

I. INTRODUCTION

Several search engines were designed to get information which is adopting keywords matching mechanism. Keywords matching are insufficient due to the retrieval of a large amount of irrelevant information, which has shortness in understanding query intentions. Ontology allows semantic analysis of input queries and a heuristic search, the expected information can be retrieved more precisely and completely that is satisfying consumer's intentions.

Generic approaches were demonstrated for the last years that intend to extract information from documents which are available on the World Wide Web or HTML page. Those previous method exploits redundancy of information to compensate for loss of precision caused by the use of domain independent extraction methods, and presents an outline in the form of a framework that is applicable in various domains.

This motivates this research to design a new algorithm consists of hybrid instances extraction algorithm and semantic matching to improve the matching results. The main aim of this research is to enhance the text matching, where a case study was introduced for the renewable energy, thus, the matching will be done between renewable energy providers and consumers' requirements.

Enhance the matching process may include enhancing the keywords matching by using semantic matching. Also the semantic matching could be improved by retrieving information that are more precise and satisfies the consumers' intentions.

This research is based on instances that help in enhancing the semantic matching results. It is difficult to extract instances from various data sources manually. Several algorithms where designed to extract instances from texts. In common, populating an existing ontology with instance information present in the natural language text provided as input to extract the instances. The heuristics are adapted to extract information from the unstructured text and for adding it as structured information to the selected ontology. This identification of the relevant ontology is critical, as it is used in identifying relevant information in the text.

Ontology should play a good role in the software life cycle. It's defined as a formal, explicit specification of a shared conceptualization. Generally ontology provides set of concepts and their interrelationships in a specific domain to facilitate understanding and automatic processing of text.

The main problem of this research is semantic matching and how it could be improved it to solve some semantic issues such as too many matching or too specific which is too small matching. Thus this research solves the problems of how the ontology could be build in the domain of renewable energy, how can the ontology serve the customer to match their requirements by using semantic matching techniques, and how the domain ontology concepts could be extracted and recognized. Thus, this paper improves the matching to be more accurate and relevant to customer's requirements. The matching will be done in terms of semantics. Semantic matching is improved by introducing the instances for match to access to relevant matching. This paper identifies the role of instances in improving the semantic matching.

This paper proposes a new way to extract instances that is based on mathematical expression and statistical analysis.

The instances are extracted features and requires more processing to get a meaningful result or a result that could be used in decision making process. The instances detection procedure will result two different sets of instances; one regarding to the user input queries and the other is regarding to the supplied renewable energy vendors. The semantic matching will be done between those two sets of instances.

The decision making process is the process of that, getting some of vendors with vendor profiles, and inputting a customer specific queries. Then, applying the proposed algorithm of this paper in order to determine which vendor is the most matches to the customer queries to supply it with renewable energy service.

In the field of renewable energy, the process of marking object of interest is related to huge size of data. So, the user is capable to mark the object in one page once, only one. Then, the marking page could grow automatically by adaptive instance detection algorithm.

Many algorithms could be adopted and used for instance detection. This comes from the fact that, there is no exact expression of instances in different applications and object properties.

## II. METHODOLOGY

This research aims to develop an algorithm that merges instances extraction and semantic matching in application of renewable energy. The general flow of the application methodology is that, a number of reputable vendors of renewable energy were selected carefully by constraints selection rules. After selection of the vendors, the application that used to select specific queries was designed. This application is specialized to control the application of the algorithm in addition to get the marked object for instances selection. The marking is being done manually by the user. The input is text, so, the marking process is the process of writing the queries itself in text file as the common marking procedures.

When the marked objects entered, the instances extraction algorithm will be evaluated and the output of that algorithm is a set of extracted instances that are near to the marked object texts. The extraction of instances will be done in two phases as illustrated in figure-1; the first phase is to extract the instances from the query input files those supplied by the user, while the second phase is to extract the instances from the vendor files. The vendor files are previously collected. That could be done because of that, the comparison will be done between the two types of instances (those extracted from the vendor files and those extracted from the query files).

The decision making process is being done in terms of semantic matching. A semantic matching algorithm design was adopted and implemented to compare the extracted instances in the prior steps of this presented system, then, to compare the results and extract the instances those have the higher

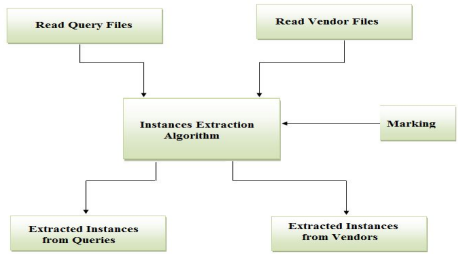

Figure-1: The main functions of instances extraction

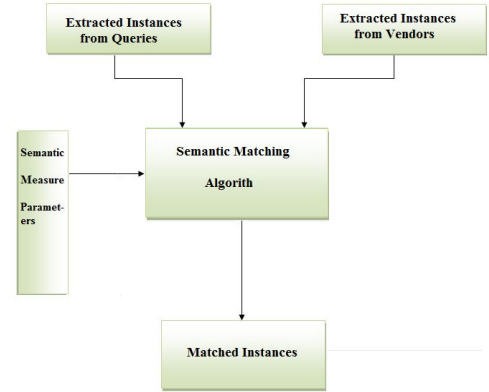

Figure-2: Block diagram of the semantic matching flowchart

match between the input customer queries and the supplied vendor files.

Both sets of extracted instances are being considered as input argument to the semantic matching algorithm. The semantic matching will give a measure for the match percentage semantically between each of the two sets of arguments. Then, the decision making part will decide either considered this measure to be a significant match or not. Figure-2 shows the semantic matching part of the system diagram.

When extracting the matched instances, one process still needed in order to take a complete decision, that is which of the vendors is better to supply the renewable energy. Hence, more than one vendor can get a match result in semantic matching, a comparison between the matched instances of all vendors should be done with respect to the matched instances and there frequencies. The percentage of match is being calculated and the vendor that get the maximum match percentage will be considered the best match one that meets the user or customer requirements in renewable energy needs.

Figure-3 illustrates the total block diagram of the presented system in this paper with detailed components, starting by user / customer input queries and supplied vendor files, and includes the instances extraction process from both, the vendor files and the input queries, in addition to the semantic algorithm and the semantically resulted instances. The decision making process after measuring the best match vendor based on the frequency of match.

Preprocessing could be made on the text data in order to make it mathematically efficient and capable to be processed. The extracted instances are subjected to semantic measure. The measurement process is the function that compares the semantic matching results of the different vendors, and computes the percentage of match, then, it selects the best match one as the reputable vendor that could provide the more suitable service to this customer based on the specific input queries.

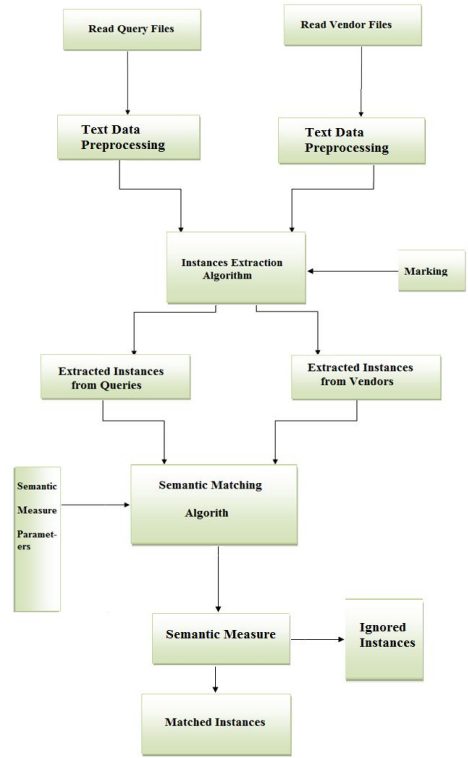

Figure-3: main block diagram of the presented paper

The best vendor that could supply the renewable energy for the specific customer could be selected based on the maximum match percentage, which is varies depending on the user input queries, and the parameter of semantic match. This is being decided depending on the decision making principal.

### 2.1 Marking of Objects

The initial process in every instance detection algorithm is to mark an object in order to compare the extracted object with them, then, to decide if that extracted object is an instance or not. The marked objects are the responsibility of the user in many cases, or the responsibility of the programmer in some cases.

For small adaptive systems, the user has to open a file that is related to object that is needed to be extracted, and then, mark all object of interest inside it. The user has to decide the object of interest by him. But in more complex systems, the marking process intended to be handled in more specialized way, because of that, the complexity of the system didn't make the user reliable in such criteria.

Since, the goal of study case of this paper is to make the renewable energy vendors selection more efficient, more reliable, and easier with respect to user, the marking process could make a big issue. That comes from the fact that, the customer is not intended to deal with complex issues in the application program like marking. So, the solution is to adapt the marking process by the programmer. Thus, the overhead of marking will be loaded to the programmer side once only. It will be loaded to the programmer once, because of that, the programmer initially should open a random renewable energy related page. Then select each object inside the page (which is a text object) and mark it manually.

The marking process by the programmer is done in a simple way. He selects the text objects those represents an object of interest with respect to renewable energy field. Then the programmer has to save the extracted instances manually in a separated text files that represents the marking page. The marked object should be saved with their frequencies.

The marking file that is created by the programmer is not a static text file, it is dynamic. It is dynamic by the means of that, the application program of the algorithm when find an object that considered as instance, and this object is not included in the marking file, it will automatically update the marking file and add the new instance as marked object with its frequency.

This contributed procedure is adaptive and make the process more reliable, but it adds more complexity on the program development side. Where, the adaptability is a complex issue in the program development strategies. But also, this procedure has three advantages; it keeps the process more simple with respect to the user, because of that, the user will not mark anymore; it puts a small overhead to the programming, because of that the programmer do that process once only, and will not repeat it; and it makes the instance marking more reliable because of the adaptability of that marking.

Table-1 shows a sample of the marked instances that is being extracted and marked manually form a random page that is related to the renewable energy. The first column of this table represents the marked objects, where the second column shows the frequency of that object in the selected page.

Table-1: sample of marked objects

| Energy Sources | 24 |
| Energy | 165 |
| Resources | 51 |
| Sun | 37 |
| Wind | 33 |

### 2.2 Instances Extraction

This paper is based - for extraction of instances - on distant supervision, where the distance is measured between each two objects to determine how they could be related to each other. The relation between objects is measured as D(obj1, obj2). The text object is an array of text elements where the single element is a character in ASCII format that represents the building block of the object. The following example in figure-4 shows the text array and its elements

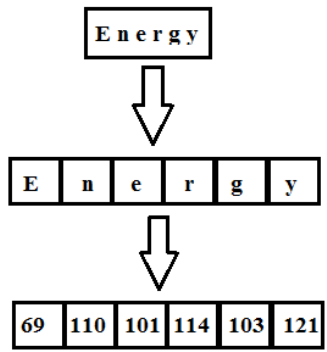

Figure-4: ASCII analysis of an object concept

The relationship between those elements could be described mathematically in different equations. The standard deviation between each two adjacent elements represents a unique structure of that object, while the variance adds more rigid structure to it. The Euclidean distance is the measure of relatedness in terms of geometrics with respect to each element of the object vector. Euclidean distance is a major measurement in different instances detection algorithm.

The relation between each two entities like (obj1 = 'Sun ' ) and (obj2 = 'Sunny') is calculated using not only Euclidean distance, but also, statistical representation of test specifications. Equation-1 shows the mathematical relationship between the two objects (R).

R=D(obj1,obj1)+ D(std(obj1,Obj2) )+ var(obj1,obj2) ... (1)

Where,

D: is the Euclidean distance (see equation-2)

std: is the standard deviation between the elements of the single object (see equation-3)

var: is the variance between the two objects, obj1, and obj2, (see equation-5)

$$d(\mathbf{p}, \mathbf{q}) = \sqrt{\sum_{i=1}^{n}(q_i - p_i)^2}. \qquad ....(2)$$

Equation-2 represents the Euclidean distance between two set of points. Where,

qi: is the ith coordinate of the set of points (q)

pi: is the ith coordinate of the set of points (p)

$$\sigma = \sqrt{\frac{1}{N}\sum_{i=1}^{N}(x_i - \mu)^2} \qquad ....(3)$$

Equation-3 shows how to calculate the standard deviation of a specific entity. The standard deviation is the most important statistical variable that enables to understand the homogenous structure of specific data set.

Where in the equation,

μ: is calculated using equation-4.

xi: is the element of the input set of specific intensity object

N: is the number of elements of the set of (objx)

$$\mu = \frac{1}{N}\sum_{i=1}^{N} x_i \qquad ....(4)$$

Equation -5 presents how to calculate the variance between two object using element set of each input, pi and xi. Where μ is defined in euqation-4

$$\mathrm{Var}(X) = \sum_{i=1}^{n} p_i \cdot (x_i - \mu)^2 = \sum_{i=1}^{n}(p_i \cdot x_i^2) - \mu^2 \qquad ....(5)$$

The calculated variable R in equation-1 is used to determine if the weight of the marked object that subjected to instances measure from the input vendor profile text file or query file with respect to the marking object that selected by manually marked objects file. If R value gets lower than specific threshold number, then, the two objects (obj1, and obj2) will be considered to be related objects.

Supposing that, obj1 is the marked input object and obj2 is the selected object from text files of vendors or queries, then, if R less than the determined threshold, then, obj2 will be considered as instance.

This paper determined R value to be 0.01, if its value that gotten by equation-1 is less than 0.01, then, obj2 will be considered to be an instance, and it will be added to the instance list to be extracted from the vendor profile. But in case of that, the selected object from vendor or query files is not exist in the manually marked objects and not similar to any one of the marked object. Then this paper assumes that, the presented algorithm should handle that object.

The handling of objects that do not exist in the marked list, are done by measurement of that object match with all other marked objects. If the matching value between the selected object and any one of the manually marked objects got high matching value, then, it will be considered to be an instance.

This procedure selects the instances and extract them from the vendors files and the query files respectively. The extracted instances will be either exists in the manually marked objects file, or either match the manually marked objects match with the extracted instance.

The dynamic objects are texts and there is no need to add another text caption to define its, and it's enough to define it by itself. By trials, the match percentage is selected to be less than 0.009. No scientific approach to calculate or estimate this value, but the trial and error only.

When instances extracted, it should be saved in structure array, that is arranged based on the different vendor profiles. Then instances could be extracted directly from vendor profile and saved in structure form to enable using it in the later part of the algorithm. While, the instances those are extracted from the query input files, are handled in the same way as the vendor files. The same algorithm for instances extraction is used.

### 2.3 Semantic Matching

Once, the instances are ready to be handled, the matching of these instances is the next step. Now there are two structure arrays of instances; the structure array of the instances that is extracted from the vendors' profiles, and the structure array of the instances those are extracted from the queries.

The semantic matching is the process that should determine the relatedness of the input queries instances to the supplied vendors' profiles. This process will helps in decision make to decide which vendor profile is more suitable for specific query of user requirements.

Because of that this paper make a case study to help the user to decide which vendor is more appropriate to supply the renewable energy to him / his, leading to this scenario in instances detection and semantic matching.

Semantic matching algorithm where adopted for this purpose. In semantic matching, each non leave node object will be removed, and all recognized entities will be used in semantic similarity measure, in order to measure its relationship with the original input queries.

The semantic similarity concept consists of a set of input entities defined as equation-6. Where in our case, Px represents a set of extracted matches those were gotten as presented in figure-2. In addition to that set of entities (P), another set of agreements should be defined to perform the semantic similarity, let say "A". The instances are needed to keep the community common ontology. The set of semantic similarity represents a set of search information, hints, or keywords. In this research paper, "A" is the input queries that are the customer marking inputs.

$$P = \{P1, ..., Pn\} \quad .... (6)$$

The description features of the set of entities (P) should be published and distinguishable at the same time, in order to enable finding an appropriate matching algorithm. Hence, the published features are the same as calculated for the input queries those used in equations (1, 2, 3, and 5) for instances extraction process, those features in this paper are considered the published features or advertises and will be used in semantic matching process.

The entities or pears (P) is not needed to broadcast itself, hence, it is well known and its specs, computations, and parameters is easily computable using the previous work in instances extraction algorithm. So, the mapping between the input entities (defining it as agreements) and the extracted features will be used instead of concepts extraction for semantic matching of instances process.

The semantic matching is the use of semantic similarity to verify if two semantic objects are similar or not, which is a special case of semantic relatedness. The semantic similarity considers a specific type of concepts relationships which is so called hypernymy / hyponymy (i.e. IS-A relationship). In contrast, the semantic relatedness is the method to find the relationship between two concepts. But it is specialized by using special relationships between the concepts.

This paper concerns to implement semantic matching between two type of extracted instances; the instances those are extracted from the customer input queries and the instances those are extracted instances from the vendors text file represents that vendor profile. Thus, this semantic matching is concerning on phrase matching.

The phrase is restatement of text that get a specific meaning in another form or in its specific form. For example let consider the following phrases concatenation

Power <=======> White

Those two phrases are not related at old to each other, but the mathematical description of them, is very quite. So, the mathematical description is not capable to perform an accurate verification or matching between different concepts if those concepts are phrases. This case illustrates why the semantic matching and processing is required. But this is not the single advantages of it in this research; the matching power itself is an important point.

The above two concepts are very far from each other, in meaning and also in its linguistic structure. But also, the two phrases contain a very view difference when applying an Euclidean distance measure, that described in equation-3. Where the Euclidean distance between them (using the equation-2) is ==> 0.2298

Also, there are more complex sentences, where the Euclidean distance measure between them is absolutely zero. For example, let consider the following two sentences

Clock <=======> Lock

The above two sentences has Euclidean distance measure between them near to zero (i.e. 0.00263). Thus, the matching between instances using statistical meanings is not reliable and may lead to misguide the designed system and thus.

Lexical comparison could be used to match two phrases, in classical way. But this will limit the matching to digital results or one to one match. No machine intelligence is available in that way. The lexical matching is normally failed in decision making process, because of that, phrases with the same meaning will be measured as un-match in lexical comparison.

The lexical comparison is the meaning of comparing two text sentences directly by string or character features. Like normal text compare function those are being used in basic programming, or even by complex text comparison techniques where the text is represented in another format like hash-table for example. For example, consider the following two words:

Sun <==> Solar

Those two words (sun and solar) are referred to same meaning in renewable energy, which is the ability of sun rays to generate an electrical power directly without a need of any environmental waste tooling. The word solar energy is the most common in engineering applications and environments, where the sun refer to the star of solar system, and it is less used in engineering terms. But the two words are the same and refers to the same reference when we use them as

Solar energy = Sun energy

Where, in lexical matching those are different and not related to each other at old. So, the need of semantic matching in such applications is more required, because of its reliability, efficiency, and the realistic output data with verity of concepts.

For those reasons, the selection of semantic matching is for reliability and efficiency in such cases, where the matching between two sentences or two textual objects is being done in terms of object semantics. This research implements semantic matching algorithm based on (WuP) semantic measure algorithm [1].

Phrase matching is required in many applications, like extraction or retrieval of information, automatic translation, copyright infringement, automatic identification processes, and other applications. This paper - as proposed - study the application of phrases matching to make a decision in the field of renewable energy regarding. This decision is related to that, determining which vendor - from a list of well known vendors - are the best match for specific user input query.

The adoption of semantic matching solves all problems that are related to text semantics, including the meaning semantics and the lexical structure semantics. It gives rigid comparison criteria to find the similarity.

The semantic similarity matching requires a metric to find if the two concepts are related to each other or not, and if yes, how much. They key difference between the concepts should be considered in those metrics, and not only the general shape, but the core meaning of the phrase will be taken into account.

To determine if the two concepts are similar semantically, or not, a measurement of match should be computed. Many measurements are available for semantics and matching. The most common measurements for similarity are Path Length, Conceptual Distance, Random, Wu-Palmer, Jiang & Conrath, Resnik, and Leacock & Chodorow.

This paper refers to Wu-Palmer method (WuP). This method calculates the similarity between two concepts using the depth of the two concepts; the depth is in UMLS with the LCS depth. Equation-7 shows how to calculate the WuP measure results, which is commonly known as score [1].

$$Score = 2x \frac{depth\ (LCS)}{(depth(s1) + depth\ (s2))} \quad \ldots (7)$$

where:

s1: is the first concept

s2: is the second concept

score: have a value between 0 and 1

The UMLS in semantic matching represent a useful relationship between two types semantically. It is originally comes from the Unified Medical Language System. It is based on categorizing all concepts those are presented in UMLS in special arrangement.

The LCS is defined as the Least Common Sub-summer. It represents the shared present between two nodes with deepest value and it is separated from the root concept. Where if the LCS goes more deeper, it means that, the similarity measure is more large.

As equation-7, the larger LCS gets larger Score in the similarity measure based on WuP approach for semantic measures. The LCS supports the knowledge based systems and make the measurements more reasonable.

Hence, the depth of LCS is always greater than zero, so, the score will not be zero in any case. If the two concepts are identical, then the score value will be one. Otherwise, it will be less than one. For example, if the two instances are the same like the following two sentences, then the WuP score result will be one. So the result is bounded by [0, 1]

Wind Speed <==> Speed of Wind

The algorithm designer should determine if the value of the score could be considered to be a match or not. Actually, there is no scientific analysis or procedure to calculate or estimate this threshold value. Instead, it is normally determined based on trial and error procedure. A set of instances where selected and the WuP score is calculated continuously for them, then, by human experience, it is easy to determine the best value that could be useful in such application. A 0.9 score is considered to be match.

Actually, trying to use value larger makes the matching process more constrained, and the matching process will not pass successfully. While chosen a lower value for the score threshold will cause the decision making process to be unreasonable. So, the correct selection of the threshold value is an important process in semantic matching and measurements. The resulted matched semantic instances need to calculate the match percentage in order to make a decision.

The percentage of match is being calculated depending on how a specific vendor scores and how many matches it achieve with each query and with all queries. So, the percentage of match is a function of semantically matched instances and its frequency.

The decision making process involves the comparison of the match results and determine the maximum value of match and the vendor that achieves that value. The decision making will be a selection of the vendor that supplies better renewable energy options, service, and other features.

III. RESULTS

The application program that was designed in order implement the algorithm that is presented in this paper aims to proof the concept of the presented algorithm. The implementation was very useful in parameter estimation of the presented algorithm; in addition to complete the process of trial and error in finding the best fit values for some equations and selected factors (e.g. R value of the instance detection algorithm, and the WuP score).

The implementation was built on MATLAB programming language, which is high level language with full mathematical and control flow abilities.

The implemented program enables the user to directly fill the requirements query in a ".txt" format file. Then, the program will read it directly from the file path. The algorithm will be able to run on the vendor files and the input queries together.

The test data set sample is consists of two arrays; the first is the vendors array which is collected from 10 vendors, while the second is the queries array which collected of 30 queries.

Figure-5 shows a sample of input query. The well established written input query is the finest extracted instances, thus, better matching results.

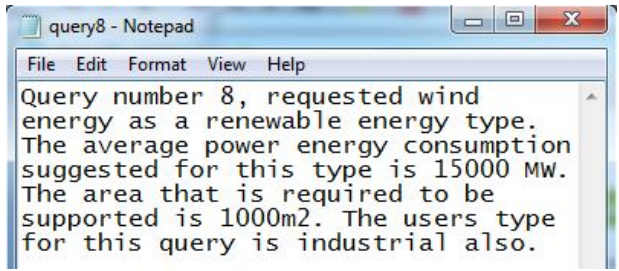

Figure-5: Input query sample

Figure-6 shows result of extracted instances of vendors files. Where the column index represents the extracted instance index and the row index represents the index of vendor which the instances extracted from its text profile.

In the same way, result of extracted instances of queries files is represented. In decision making process, the greatest match percentage vendor is the more appropriate vendor supply the customer with that input query.

Figure-6: Example of extracted instances result that is extracted from vendors

Table-2 shows comparison between the presented algorithm and related works that is used for semantic matching and/or instances detection. The comparison was made in terms of both, the concept of the algorithm and the result accuracy.

Table-2: comparison between related and presented algorithm

| Algorithm | Pengbin, 2009 | Sui, 2012 | Bhavana, 2012 | Presented |
|---|---|---|---|---|
| Main Idea | Semantic matching for web pages concepts. The semantic measure is based on semantic distance | Synchronous attributes based instances extraction. the extraction is contextual similarity based and association between concepts | This paper extracts instances pairs from HTML file, it's based on HTML clustering tables using Hearst patterns | This paper is related to two subjects, the first is to extract instances from vendor and input query, and the second is to find the best match between queries and vendors using semantic matching |
| Application field | Semantic web services | Web services | HTML corpus pages | Renewable energy field |
| Marking procedure | No marking | By programmer | Automatic marking | The user marks manually one random page, then the marking is updated automatically |
| Accuracy | 78% | 69% | 84.5% | 91% |

## IV. Conclusions

The results show that, the presented algorithm accuracy is much higher than the related works. This comes from the contributed hybrid technique which merges both, the instances extraction and the semantic matching. Where the instances exctraction work as feature extraction algorithm to minimize the data set before implementing the semantic mathchign algorithm.

The better matching result depends on the instances extraction algorithm, matching algorithm, and the ability of the input query to determine the exact query demand for the target data set (i.e. renewable energy this case study). Also, it depends on the vendor itself and the ability to supply specific renewable energy features or services to the customers. In fact, if the user queries was not correct as possible as enough to get the scope of the query requirement application, the match percentage will be not complete depending on the query. So, it is important to select the correct reasonable query and expressed vendor data. This differs from one case to another, and depends on the case of vendors, consumer specs, location, and sure, the cost of that supplied system in the determined case.

The merge of instances extraction and the semantic matching increase the accuracy of match up to 91% which is much higher than the compared related papers / algorithm. The efficiency and reliability of the hybrid techniques was very useful and could be improved more over to get higher accuracy percentage in many researches.